\documentclass{emulateapj}
\usepackage{enumitem}

\newcommand{\kms}{\ifmmode {\rm km~s}^{-1} \else km~s$^{-1}$\fi}
\newcommand{\Msun}{\ifmmode {\rm M}_{\odot} \else M$_{\odot}$\fi}
\newcommand{\Lsun}{\ifmmode {\rm L}_{\odot} \else L$_{\odot}$\fi}
\newcommand{\qo}{\ifmmode q_{\rm o} \else $q_{\rm o}$\fi}
\newcommand{\Ho}{\ifmmode H_{\rm o} \else $H_{\rm o}$\fi}
\newcommand{\ho}{\ifmmode h_{\rm o} \else $h_{\rm o}$\fi}

\newcommand{\vFWHM}{\ifmmode v_{\mbox{\tiny FWHM}} \else
                    $v_{\mbox{\tiny FWHM}}$\fi}
\newcommand{\CCF}{\ifmmode F_{\it CCF} \else $F_{\it CCF}$\fi}
\newcommand{\ACF}{\ifmmode F_{\it ACF} \else $F_{\it ACF}$\fi}
\newcommand{\Halpha}{\ifmmode {\rm H}\alpha \else H$\alpha$\fi}
\newcommand{\Hbeta}{\ifmmode {\rm H}\beta \else H$\beta$\fi}
\newcommand{\Hgamma}{\ifmmode {\rm H}\gamma \else H$\gamma$\fi}
\newcommand{\Hdelta}{\ifmmode {\rm H}\delta \else H$\delta$\fi}
\newcommand{\Lya}{\ifmmode {\rm Ly}\alpha \else Ly$\alpha$\fi}
\newcommand{\Lyb}{\ifmmode {\rm Ly}\beta \else Ly$\beta$\fi}
\newcommand{\HeI}{\ifmmode {\rm He}\,{\sc i}\,\lambda5876 \else 
	          He\,{\sc i}\,$\lambda5876$\fi}
\newcommand{\HeII}{\ifmmode {\rm He}\,{\sc ii}\,\lambda4686 \else 
	           He\,{\sc ii}\,$\lambda4686$\fi}

\newcommand{\heii}{He\,{\sc ii}}

\newcommand{\ciii}{\ifmmode {\rm C}\,{\sc iii} \else C\,{\sc iii}\fi}
\newcommand{\civ}{\ifmmode {\rm C}\,{\sc iv} \else C\,{\sc iv}\fi}
\newcommand{\CIV}{\ifmmode {\rm C}\,{\sc iv}\,\lambda1549 \else 
	           C\,{\sc iv}\,$\lambda1549$\fi}

\newcommand{\oii}{O\,{\sc ii}}
\newcommand{\oiii}{O\,{\sc iii}}
\newcommand{\ob}{[O\,{\sc iii}]\,$\lambda \lambda 4959,5007$}
\newcommand{\oiv}{O\,{\sc iv}}

\newcommand{\mgii}{Mg\,{\sc ii}}

\newcommand{\siiv}{Si\,{\sc iv}}

\newcommand{\caii}{Ca\,{\sc ii}}

\shorttitle{Quasar-Diversity Biases in BOSS Redshifts}
\shortauthors{}

\received{}

\begin{document}

\title{The Sloan Digital Sky Survey Reverberation Mapping Project: Biases in \lowercase{$z>1.46$} Redshifts due to Quasar Diversity}

\author{ K.~D.~Denney\altaffilmark{1,2,3}, Keith Horne\altaffilmark{4}, W.~N.~Brandt\altaffilmark{5,6,7}, C.~J.~Grier\altaffilmark{5,6}, Luis~C.~Ho\altaffilmark{8,9}, B.~M.~Peterson\altaffilmark{1,2},  J.~R.~Trump\altaffilmark{5,11}, J.~Ge\altaffilmark{10}}

\altaffiltext{1}{Department of Astronomy, 
		The Ohio State University, 
		140 West 18th Avenue, 
		Columbus, OH 43210, USA;
		denney@astronomy.ohio-state.edu}
		
\altaffiltext{2}{Center for Cosmology and AstroParticle Physics, 
                 The Ohio State University,
		 191 West Woodruff Avenue, 
		 Columbus, OH 43210, USA}

\altaffiltext{3}{NSF Astronomy \& Astrophysics Postdoctoral Fellow}

\altaffiltext{4}{SUPA Physics/Astronomy, 
		    Univ. of St. Andrews, 
		    St. Andrews KY16 9SS, Scotland, UK}

\altaffiltext{5}{Department of Astronomy \& Astrophysics, 
                      525 Davey Lab, The Pennsylvania State University, 
                      University Park, PA 16802, USA}
			
\altaffiltext{6}{Institute for Gravitation and the Cosmos, 
                      The Pennsylvania State University, 
                      University Park, PA 16802, USA}
			
\altaffiltext{7}{Department of Physics, 
                      104 Davey Lab, The Pennsylvania State University, 
                      University Park, PA 16802, USA}

\altaffiltext{8}{Kavli Institute for Astronomy and Astrophysics, 
                     Peking University, 
                     Beijing 100871, China}

\altaffiltext{9}{Department of Astronomy, 
                      School of Physics, 
                      Peking University, 
                      Beijing 100871, China}

\altaffiltext{10}{Astronomy Department
			University of Florida
			211 Bryant Space Science Center
			P.O. Box 112055
			Gainesville, FL 32611-2055, USA}

\altaffiltext{11}{Hubble Fellow}


\begin{abstract}
  We use the coadded spectra of 32 epochs of Sloan Digital Sky Survey (SDSS) Reverberation Mapping Project observations of 482 quasars with $z>1.46$ to highlight systematic biases in the SDSS- and BOSS-pipeline redshifts due to the natural diversity of quasar properties. We investigate the characteristics of this bias by comparing the BOSS-pipeline redshifts to an estimate from the centroid of \heii\,$\lambda$1640.  \heii\ has a low equivalent width but is often well-defined in high-S/N spectra, does not suffer from self-absorption, and has a narrow component that, when present (the case for about half of our sources), produces a redshift estimate that, on average, is consistent with that determined from [\oii] to within 1$\sigma$ of the quadrature sum of the \heii\ and [\oii] centroid measurement uncertainties.  The large redshift differences of $\sim$1000 \kms, on average, between the BOSS-pipeline and \heii-centroid redshifts suggest there are significant biases in a portion of BOSS quasar redshift measurements.  Adopting the \heii-based redshifts shows that \civ\ does not exhibit a ubiquitous blueshift for all quasars, given the precision probed by our measurements.  Instead, we find a distribution of \civ\ centroid blueshifts across our sample, with a dynamic range that (i) is wider than that previously reported for this line, and (ii) spans \civ\ centroids from those consistent with the systemic redshift to those with significant blueshifts of thousands of kilometers per second.  These results have significant implications for measurement and use of high-redshift quasar properties and redshifts and studies based thereon.

\end{abstract}




\section{INTRODUCTION}

Mapping the location of stars and galaxies as a function of distance --- or redshift --- helps us understand not only the contents of the Universe but also its structure and evolution and the physical principles shaping what we observe. Quasars, or active galactic nuclei (AGN; used synonymously in this work), are arguably the most useful extra-galactic source for mapping the Universe at high redshift.  These accreting super-massive black holes (BHs) can outshine their host galaxies by several orders of magnitude and are thus observable at much greater distances than their quiescent counterparts.  Moreover, quasar spectra are characterized by the presence of high equivalent width (EW) emission lines distributed across UV to NIR wavelengths.  These emission lines can be identified and redshifts determined even with relatively low S/N, resource-economic, ``survey-quality" spectra. Several large surveys have been obtaining large numbers of quasar spectra for cosmological studies, such as measurements of baryon acoustic oscillations \citep[e.g.,][]{Busca13, Delubac15}.  The Sloan Digital Sky Survey (SDSS), alone, with programs such as the Baryon Oscillation Spectroscopic Survey (BOSS), has spectroscopically confirmed $\sim$370,000 quasars \citep{York00, Schneider10, Dawson13, Paris14}.  

These quasar redshifts are invaluable for studies on both large and small scales, both for studies directly related to the quasars and for those reliant upon the intervening absorption. However, making accurate redshift measurements of quasars, particularly at high redshifts, is surprisingly difficult --- an issue that is, perhaps, not broadly known or appreciated outside the direct quasar physics community.  Quasar spectra are a blended superposition of many emission and absorption components that arise from physically distinct sources at different distances from the BH. Components include the thermal continuum from the accretion disk, narrow and broad emission lines from the narrow-line region (NLR) and broad-line region (BLR), respectively, intrinsic and intervening absorption lines, and host-galaxy starlight.  Of these emission- and absorption-line components, some are better-suited for redshift determinations than others.  

Associated quasar absorption lines are usually attributed to outflows from the nucleus and are not expected to lie at the systemic redshift. Host-galaxy stellar absorption features, on the other hand, are the most robust measure of the galaxy's redshift.  Unfortunately, these absorption lines are generally masked by the luminous quasar contributions.  Even if observable, rest-frame optical lines, such as \caii\ H \& K lines at $\lambda\lambda$ 3969, 3934, are inaccessible in optical spectra of high-redshift quasars. 

In the absence of reliable absorption lines, quasar redshifts are best determined from narrow emission lines that arise from the NLR or the host galaxy.  Of particular interest are lines due to forbidden transitions that cannot arise in the high-density and high-temperature BLR environment, and so are not blended with a broad-line component or severely susceptible to dynamics dominated by the nuclear activity. The narrow lines are still sometimes observed to have small blue-shifts ($\sim$10$-$100\,\kms) compared to host-galaxy absorption lines, and this effect may have a luminosity dependence \citep[see, e.g.,][]{Woo16, Shen16b}.  Unfortunately, using high-EW, isolated, narrow forbidden lines also becomes difficult using optical spectra of high-redshift quasars because rest-frame UV forbidden emission lines have much smaller EWs, and, at the highest redshifts ($z\gtrsim$\,3), the only NLR emission lines visible in optical spectra are due to permitted transitions that are blended with BLR emission, have relatively low EW, and are often resonance transitions susceptible to self-absorption, e.g., \Lya, \CIV, or \heii\, $\lambda1640$.

If absorption and narrow emission lines cannot be used for redshifts, the broad emission lines are used.  This is most often required at high redshifts, where the other methods are no longer suitable.  Difficulties in measuring redshifts based on these lines arise first from fundamental difficulties with either the specific transition or data quality (or both).  First, the strong UV lines (\Lya, \civ, \ciii], and \mgii) are all either resonance lines susceptible to self-absorption, and/or are heavily blended with other species or within a multiplet. The line ratios within these blends and multiplets depend on physical properties of the nuclear environment, such as optical depth, density, and incident ionizing radiation \citep[e.g.,][]{Baldwin95, Korista97, Casebeer06}. 

SDSS-I/II-pipeline redshifts\footnote{see http://classic.sdss.org/dr7/algorithms/redshift\_type.html} (hereafter SDSS-pipeline) are based on either emission-line matching or cross correlation with the \citet{VandenBerk01} composite quasar spectrum.  While using a composite spectrum formed from many thousands of survey-quality quasar spectra will provide a very high S/N template, it does not account for the intrinsic diversity in the physical structure, environment, and spectral energy distributions (SED) among quasars.  It will thus create biases in the redshifts of objects with properties different than the average properties of the quasars used in its creation. Of particular importance is understanding (i) the physical properties that modify the structure of the emission lines, and (ii) on what observable variables those physical properties depend.  One challenge is that the SDSS pipeline assumes only a single wavelength for many multiplets and/or highly blended transitions in the UV for its emission-line matching and line identification.  This assumption combined with diversity in the spectral structure of individual quasars (i.e., differences in multiplet ratios for those with relatively wide velocity separation) could easily contribute systematic uncertainties in the determined redshifts beyond those typically quoted based on [\oiii] or \mgii\ \citep[$\sim$50$-$300 \kms;][]{Shen11}.

Known velocity shifts between different quasar emission lines are another challenge that contributes to biases in redshifts \citep[e.g., see][and discussion and references therein]{Shen16b}.  Several studies have already explored biases in the SDSS-pipeline redshifts (e.g., \citealp{Hewett10}, hereafter HW10). While they still utilize cross-correlation with a master template, HW10 improve the SDSS redshift estimates of quasars by building a redshift `ladder' as a function of increasing redshift, since the redshifts of more nearby quasars can be more accurately determined from host-galaxy stellar features and strong forbidden, narrow emission lines.  HW10 and \citet{Shen16b} find trends in line-to-line velocity shifts with quasar luminosity.  While luminosity is an easily measured observable, more in-depth analyses of the spectral diversity of quasars from eigenvector analysis \citep[cf.][]{Boroson92} suggest that the largest source of emission-line diversity in quasars, dubbed ``Eigenvector 1" (EV1), is more likely related to accretion rate and the quasar spectral energy distribution (SED).  EV1 analysis shows that the relative strengths of different emission lines and the velocity shifts between lines seem to be well-correlated \citep[e.g.,][]{BaskinLaor05}. In flux-limited surveys, luminosity can be a reasonable proxy for accretion rate, potentially leading to the luminosity correlation found by HW10 and others.  This connection is a concern for redshifts based on composite quasar spectra formed from flux-limited samples, as the composite will be weighted toward the spectral properties associated with the relatively higher luminosity quasars.

Redshift estimates from the BOSS pipeline \citep{Bolton12}, which we utilize in this work, use an eigenvector principle component analysis (PCA) method that is more sophisticated than the previous SDSS-pipeline cross-correlation redshifts. Cross-correlation and PCA analysis should both be more robust against biases due to small, intrinsic velocity shifts between lines \citep{Shen16b} than using individual emission lines to determine the redshift because these methods average over all shifts.  Additionally, the PCA-based redshifts should be even more robust than cross correlation with a single composite spectrum because a template built for each quasar is not as susceptible to the ``averaging" biases from using a single quasar template.  However, the BOSS PCA templates are built using training-set spectra with redshifts determined from the original SDSS-method.  As such, while the BOSS PCA method is more sophisticated overall (see also \citealp{Dawson15} for continued improvements), redshift biases \citep[see, e.g.,][]{FontRibera13, Paris14} may have propagated into the templates due to redshift inaccuracies in the training set spectra. 

A particular problem associated with accurate quasar redshift estimates is whether BH masses can be well-estimated using the \civ\ emission line.  In particular, the apparent blueshifts in the \civ\ line have been used as evidence that the \civ\ velocity widths are indicative of non-virial motions and therefore ill-suited for virial BH mass calculations \citep[see also][]{BaskinLaor05, Sulentic07, Netzer07, YShen08, Shen12}. However, since radiation-driven winds generally have velocities comparable to the BH escape velocity \citep{Cassinelli73}, they might still be quite reasonable virial estimates in this scenario.  In any case, the inference of significant and/or generally ubiquitous blueshifts in \civ\ emission depends on the reliability of the redshifts. 

Here, we investigate the possibility of systematic errors in the redshifts of quasars using BOSS spectra taken as part of the SDSS-RM Project.  We use two narrow emission lines observable in optical spectra of intermediate- and high-redshift quasars. Our investigation is laid out as follows:  In Section~\ref{S_Data} we present the SDSS Reverberation Mapping Project data that we use for this investigation. Section~\ref{S_SpecAnal} describes our analysis of these data for measuring the redshifts from \heii\,$\lambda$1640 and the [\oii]\,$\lambda$3727 doublet, and in Section~\ref{S_Discussion} we discuss our results.  Final remarks on our results are made in Section~\ref{S_Conclusions}. 

\section{Data \civ}
\label{S_Data}

The SDSS Reverberation Mapping Project (SDSS-RM) is spectroscopically monitoring broad-line quasars in a single 7 deg$^2$ field (the CFHT-LS W3 field) with the SDSS telescope's \citep{Gunn06} BOSS spectrograph \citep{Smee13}.  Here, we utilize the data obtained from the SDSS-III \citep{Eisenstein11} ancillary program between 2014 January and July.  Covering redshifts over the range $0.1 < z < 4.5$, the SDSS-RM sample consists of 849 quasars with a flux limit of $i_{\rm psf} = 21.7$ mag.  Each of the 32 epochs of observations was a $\sim$2 hr exposure taken during dark/grey time, with an average cadence of $\sim$4 days over this $\sim$6-month period.  The technical overview of this program is provided by \citet{Shen15b}.  

\begin{figure}
\epsscale{1.2}
\plotone{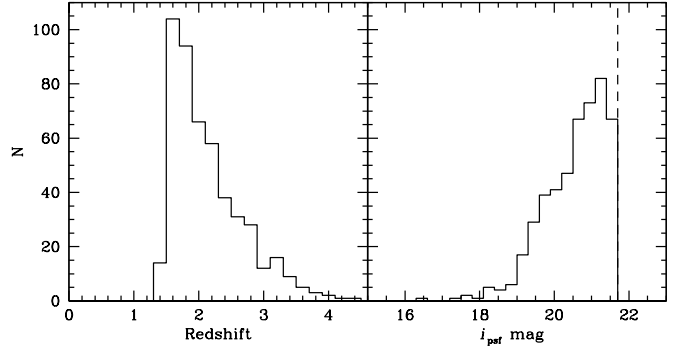}

\caption{Distribution of redshifts and $i_{\rm psf}$ magnitudes for the sample of 482 $z>1.46$ SDSS-RM quasars.  The vertical dashed line in the right panel shows the magnitude limit for the SDSS-RM sample.} 

\label{F_physprop}
\end{figure}

Here we use the subsample used by \citet{Denney16a} to study the properties of the \CIV\ emission-line region. The sample consists only of quasars with $z>1.46$, where objects were removed that obstructed the \civ\ analysis, such as a few with very low \civ\ equivalent width (EW) and/or several additional objects with broad absorption lines \citep[see][for further sample selection details]{Denney16a}. The final sample consists of 482 sources, and for this analysis, we only use the high-S/N ``coadded" spectrum, made by combining all good epochs using the latest BOSS spectroscopic pipeline {\it idlspec2d} \citep[see][and Schlegel et al., in prep.]{Shen15b}.  Most (405) quasars have all 32 epochs of spectra included in the coadd, and only 9 have more than 3 (10\%) epochs discarded. Figure~\ref{F_physprop}, reproduced from \citet{Denney16a}, shows the redshift and $i_{\rm psf}$ magnitude distribution of our sample.

\section{Spectroscopic Analysis: Measuring Emission-line-based Redshifts}
\label{S_SpecAnal}

The redshifts for our sample, as determined from the BOSS pipeline \citep{Bolton12}, are likely already a moderate improvement over typical BOSS-pipeline redshifts because we are able to use the coadded spectra \citep[see][]{Shen15b}, as opposed to the individual, lower-S/N single-epoch spectra (although \citealt{Denney16a} and \citealt{Shen16b} find that line centroid measurements, and thus emission-line cross correlations, are relatively more robust in the presence of spectral noise than other measured emission-line properties).  However, the real advantage afforded by this relatively higher-S/N sample is to enable the analysis of low-EW narrow emission lines, not otherwise reliably detected in low-S/N spectra, to study the effects of intrinsic quasar diversity on the redshift determinations.  We compare redshifts determined using the \heii\,$\lambda1640$ emission line, which {\it is} susceptible to spectroscopic diversity and shifts in its observed line properties \citep[likely related to EV1 effects; see, e.g.,][hereafter R11]{Richards11}, and [\oii]\,$\lambda$3727, which, presumably, is {\it not}.

\heii\ is not a resonance line and therefore not susceptible to self-absorption, and intervening absorption affects only a small percentage of objects.  It is also a more isolated transition than many UV lines, so blending is less of a concern. This makes it a more favorable line for studying the effects of quasar diversity on redshift determinations. However, there are two potential problems: First, while we would ideally isolate and study the NLR component of \heii, it is not always observed, presumably due to the same effects we are trying to study.  Furthermore, while the BLR component of \heii\ typically has a low EW, when the NLR component is absent, any remaining \heii\ emission appears significantly blueshifted and blended with (or part of) the \civ\ red shelf.\footnote{An emission feature at $\sim$1600\AA\ that also varies in strength among quasars and has yet to be identified as uniquely due to any specific ionic species or blend \citep[see, e.g][for discussions and further references]{Laor94, Marziani96, Fine10, Assef11}.}  Second, \heii\ is a high-ionization line, so an argument can be made that even the narrow component of this line arises, at least partially, in an outflow, as is suggested by the blueshifts of other high-ionization broad lines \citep[e.g.,][]{Wills93b, Sulentic95, Murray95, BaskinLaor05, Richards11, Denney12}.  

[\oii], on the other hand, is a forbidden narrow emission line, so it is emitted predominantly from the extended NLR that is not as susceptible to kinematic effects and diversity regulating the spectral properties of emission components arising deep within the potential of the BH.  [\oii] can also be emitted throughout the host galaxy, which will also not be affected by the AGN environment and will lead to even more robust systemic redshift determinations. \citet{Shen16b} find a tight correspondence between the peak of the [\oii] emission and stellar absorption lines for the lower-redshift subset of the SDSS-RM sample, with a systematic shift of only 8\,\kms\ and an intrinsic scatter of 46\,\kms.  Consequently, the redshifts determined by [\oii] for the present sample serve, in effect, as our control.  While [\oii] is the shortest-wavelength unblended forbidden emission line that is relatively-strong and still present in quasar spectra for intermediate-to-high redshifts, it still falls beyond the BOSS wavelength coverage for $z\gtrsim$\,1.78, so only 154 of our 482 quasars have [\oii] present in their spectra.  Thus, while it is reliable, it is not very applicable to high-$z$ quasars.

\subsection{Direct Determination of the {\rm \heii} Centroid} 
\label{S_Analredshifts}

Because of the diversity in the observed properties of the \heii\ line across our sample, it is not as straight-forward to perform automated emission-line fitting \citep[e.g.,][]{Shen11} and still control for the optimal number and relative shift of fit components.  We calculate the \heii\ centroid directly from the coadded spectra (i.e., not from functional-form fits to the \heii\ line) based on interactively selected wavelength boundaries.  We attempted to choose these boundaries at approximately the half maximum flux level on either side of the \heii\ peak --- selected by eye, and we use this centroid to define our ``\heii-based" redshifts.  Since the intrinsic strength of the \heii\ line, especially the NLR component, varies between objects, the boundary selection is often quite subjective. In some objects, no \heii\ line is visible at all, even for very high S/N spectra, and in others, noise or absorption obscures the peak.  In these cases, we also use other nearby emission features, such as the 1400\AA\ feature (a blend of \siiv\ and \oiv]) and the \oiii] 1663\AA\ lines, or even absorption lines, which have a high probability of being at or very near the systemic redshift (\citealt{Nestor08, Bowler14}; Allen \& Hewett, in preparation), to help inform the choice of the wavelength range that provides the most reasonable redshift through visual inspection under these circumstances.  When we could discern no reasonable criteria for measuring the \heii\ centroid, the BOSS redshift was kept.  Because of these difficulties, we assigned \heii\ redshift quality (Q) flags to each object, defined by the following criteria:  
\begin{enumerate}[itemsep=0mm]
\setcounter{enumi}{-1}
\item Very weak, or very broad, apparently blueshifted BLR component (compared to other features such as the 1400\AA\ feature and \oiii]\,$\lambda$1663 doublet) with no NLR component, or no obvious line;
\item Clean, narrow peak;
\item Noisy, but a relatively narrow peak is still clear;
\item Broader line, but a reliably determined peak;
\item Broader line, and the peak cannot be reliably contributed to a narrow component;
\item Noisy spectrum, where the peak/centroid is uncertain or possibly contaminated by noise or intervening absorption.
\end{enumerate}

\begin{figure*}
\epsscale{1.1}
\plotone{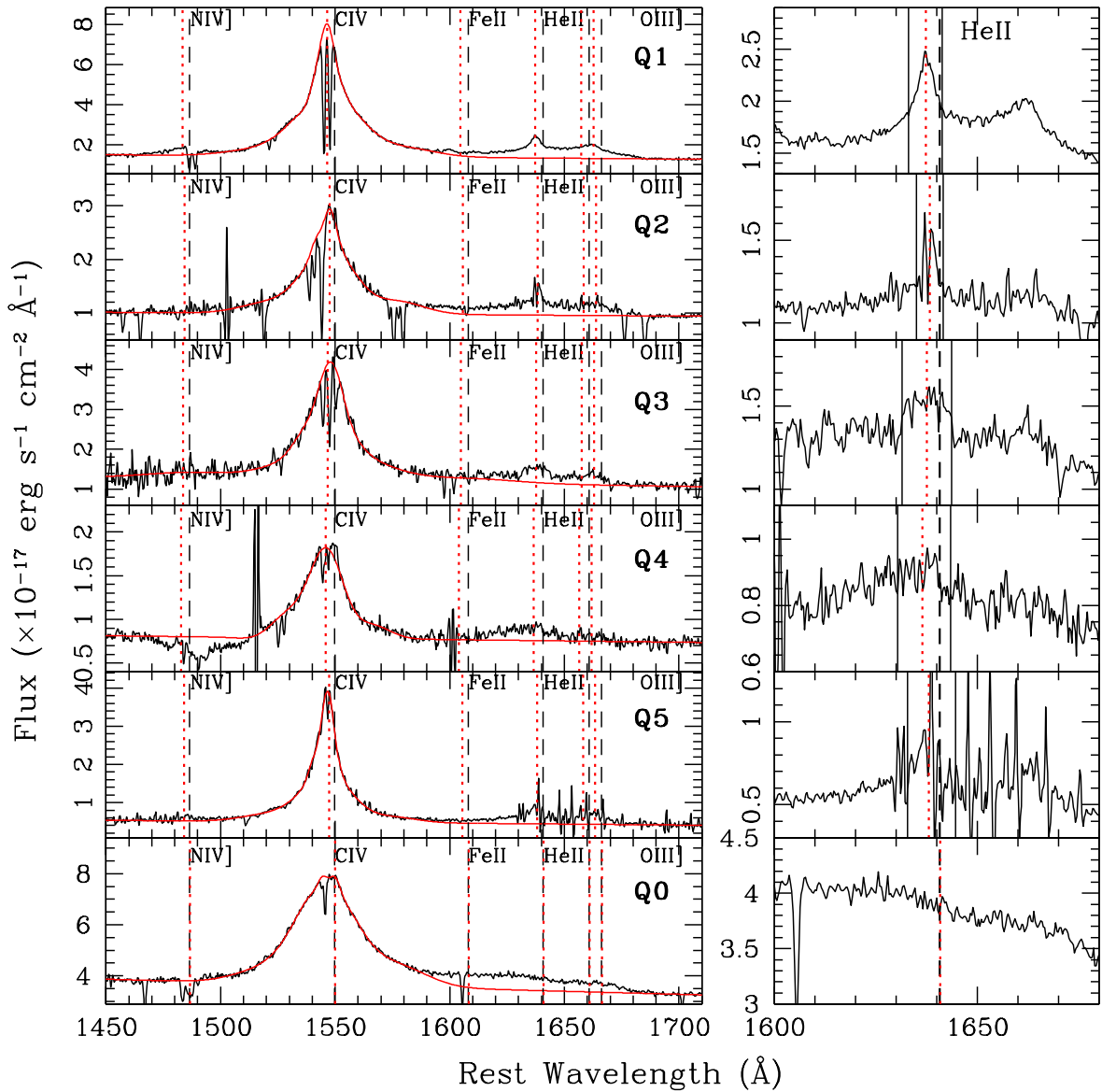}

\caption{Example spectra for each of the six redshift quality categories described in Section \ref{S_Analredshifts}.  The left panels cover the \civ\ through \oiii] wavelength region, where the original spectrum is in black and the best-fit Gauss-Hermite polynomial model for the \civ\ profile from \citet{Denney16a} is in red.  The quality (Q) category is given in the top right corner of each panel. The x-axis of each panel has been de-redshifted by the BOSS-pipeline redshift and the vertical black dashed lines show the expected location of the labeled emission lines based on the BOSS-pipeline redshift.  The red dotted vertical lines show these expected locations based on our \heii-based redshift.  The right panels show the same object as each respective left panel, only zoomed-in to the \heii\ emission line, with the expected positions of \heii\ shown again by the same vertical lines as in the right panels.  The solid black vertical lines show the ``by-eye" roughly selected boundaries for calculating each \heii\ centroid. Note that the Q=0 object does not have boundaries marked because no discernible \heii\ was visible, and so we kept the BOSS-pipeline redshift for this object.} 

\label{F_redshiftquality}
\end{figure*}

We take the redshift determinations from Q = 1, 2, and 3 objects to be relatively more robust measurements of the systemic redshift, as a strong narrow component should not be significantly blueshifted, with Q=3 the least robust due to the broader \heii\ line. Q = 4, 5, and 0 are less robust due to data-related issues, intrinsic quasar properties, or both.  Due to the overall subjectivity of these measurements, they are not to be taken as high precision redshifts, but they nonetheless provide sufficient evidence to support our investigation.  We further test the robustness of these measurements against an independent measure of the \heii\ line center in the next section.  Coincidentally, the sample is roughly evenly divided, with 237 objects flagged with Q = 1, 2, or 3, and 245 objects flagged with Q = 4, 5, or 0.  Figure~\ref{F_redshiftquality} shows an example spectrum for each of the six quality categories.

\subsection{Automated (PrepSpec) {\rm \heii} and {\rm [\oii]} Redshifts}

To form a control sample of redshifts based on a line not susceptible to EV1 effects, we use the ``PrepSpec'' analysis of this sample. PrepSpec is a reverberation mapping spectral preparation and analysis software written by one of us (KH) and applied to the SDSS-RM Project sample \citep[see][for details]{Shen15b}.  As part of the PrepSpec analysis, all BLR, NLR, and host-galaxy emission lines visible in each spectrum are modeled.  The PrepSpec output includes a measure of the modeled emission-line centers and velocity offsets with respect to the input redshift values.  While still dependent on automated modeling of the spectrum, this method provides an independent comparison for the \heii\ line center, uncertainties on the measurements, and a homogeneous methodology for comparing to [\oii] -- our control, forbidden narrow line.  We use the Prepspec \heii\ and [\oii] emission-line velocity offsets determined with respect to our \heii-based redshifts used as the input redshift. We only focus on the \heii\ and [\oii]\, $\lambda$\,3727 emission lines as a means to compare with our own \heii-based redshifts and to address the effects of quasar diversity on redshift \citep[see][for a comprehensive analysis of relative emission-line velocity shifts for this sample]{Shen16b}.  Uncertainties in the measured line centers are determined from Monte Carlo simulations.  The median 1$\sigma$ uncertainties in the PrepSpec \heii\ and [\oii] velocities are 481 \kms\ and 253 \kms, respectively.

\subsection{{\rm \civ} Blueshift Measurements}

We use the analysis of the \civ\ emission-line region by \citet{Denney16a} to measure the \civ\ blueshift with respect to both the BOSS-pipeline redshifts and our \heii-based redshifts.  We utilize the \civ\ emission-line peak wavelength measured from the Gauss--Hermite (GH) profile fits.  The uncertainties in the centroid measurements are estimated for each object from Monte Carlo simulations that create 500 flux-resampled spectra on which the measurements are repeated.  The median 1$\sigma$ \civ\ centroid uncertainty for this full sample is 183\,\kms\ \citep[see][]{Denney16a}. To be consistent with a similar study presented by R11, we calculate our \civ\ blueshifts with respect to the estimated systemic redshift such that increasing blueshifts are larger, positive velocities.

\section{Discussion}
\label{S_Discussion}

\subsection{Redshift Differences Between Independent Analyses of the SDSS-RM Sample}
\label{S_comparezSDSSRM}

\begin{figure*}
\epsscale{1.2}
\plotone{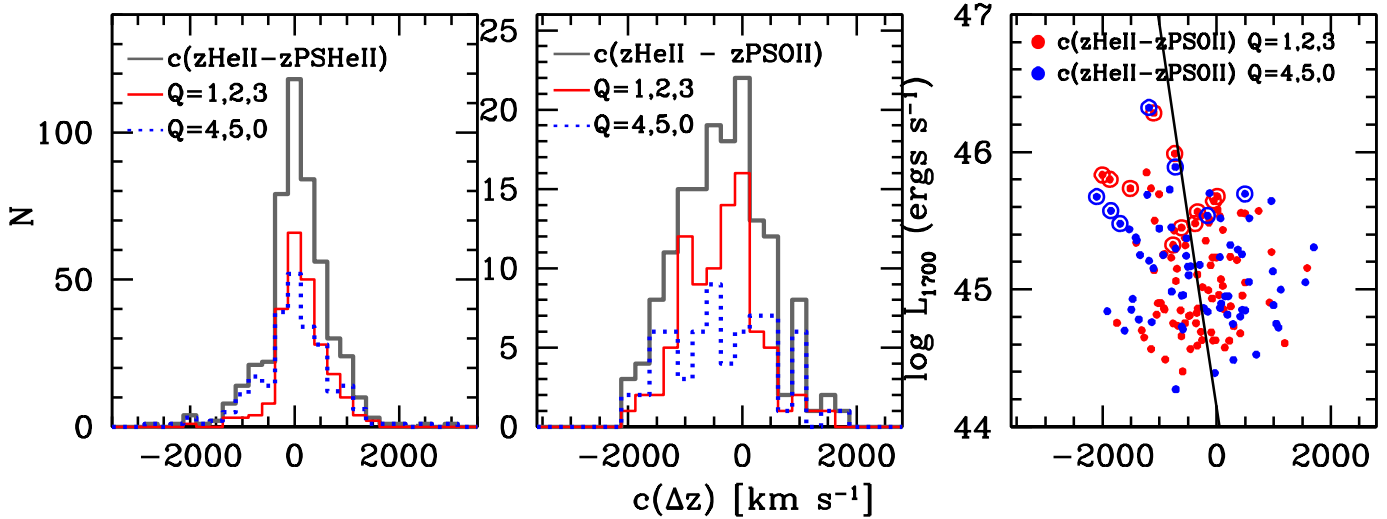}

\caption{The left (middle) panel shows redshift differences between our \heii\ redshifts and those determined for \heii\ ([\oii]) from an independent analysis using PrepSpec. The histogram colors are denoted in the legends.  The right panel shows the luminosity-dependence of the velocity shifts between \heii\ and [\oii].  The solid black curve is a simple linear least squares fit to the data. The points surrounded by larger open circles are sources that are also in the HW10 sample (see Section~\ref{S_HWcompare}).} 

\label{F_otherzcompare}
\end{figure*}

Figure~\ref{F_otherzcompare} and Table~1 show the differences between our \heii-based redshifts and those inferred from PrepSpec for \heii\ (left) and [\oii] for the 154 objects that have this measurement (middle). The PrepSpec \heii\ redshifts are consistent with our simple, direct approach, given the uncertainties.  The median difference between our \heii\ centroid measurements and the Prepspec \heii\ line centers is 54\,\kms\, which only $\sim$10\% of the median Prepspec \heii\ measurement uncertainty.  Bootstrap Monte Carlo simulations that draw randomly from the distribution shown in the left panel of Figure~\ref{F_otherzcompare} 10,000 times, with no weighting for measurements drawn multiple times, estimate a 1$\sigma$ (3$\sigma$) uncertainty in this median systematic shift of $\pm$22\,\kms\ ($\pm$62\,\kms), further supporting the consistency between our rough, visually-determined centroid-based \heii\ line centers and those determined through automated fitting.

The comparison with [\oii] (middle panel of Figure~\ref{F_otherzcompare}) shows a relatively large scatter of 780\,\kms, although a small systematic blueshift of \heii\ relative to [\oii] is evident. We find a distribution median of 348\,\kms\ and a bootstrap 1$\sigma$ (3$\sigma$) uncertainty on the median of 104\,\kms\ (242\,\kms).  The scatter is defined by the half of the 16$-$84\% inter-percentile range (HIPR) of the distribution that would correspond to 1$\sigma$ if the distribution were Gaussian, and the uncertainty in the median is again based on 10,000 trials.  Nonetheless, this systematic shift is still within the 1$\sigma$ quadrature summed \heii\ and [\oii] Prepspec measurement uncertainties. While it is difficult to assign uncertainties to our direct \heii\ redshift estimates based on the interactive centroid calculation, the systematic shift in the \heii--[\oii] redshift distribution is also within the observed scatter of the comparison between the interactive and the Prepspec \heii\ line centers, which as show above, were statistically consistent at the 3\,$sigma$ level. This suggests that the typical \heii--[\oii] line center shift is also likely within or near the order of the uncertainties of our direct \heii\ centroid measurement method.

However, since quasar diversity effects are expected to be imprinted on \heii\ but not [\oii], we also measured the systematic shift after splitting the sample by Quality rating.  We find relatively larger systematic shifts and more scatter for the more uncertain Q=4, 5, 0 \heii\ redshifts than for Q=1, 2, 3 (see Table~1). Despite the contributions of noise to the Q=4, 5, 0 distribution, this is also as expected if the primary source of the observed velocity shift is EV1 effects within the diversity of quasar spectra, from which we expect the strength of narrow \heii\ to anti-correlate with its the broad+narrow emission-line width and the velocity shift away from systemic.  This physical effect is likely additionally exacerbated within the Q=4, 5, 0 subsample by the uncertainty in our inability to isolate the \heii\ narrow component when it is weak or absent.

\citet{Shen16b} find similar but smaller relative velocity shifts, with a median \heii\ shift of $-175$\,\kms, i.e., blueward, with respect to [\oii] using yet another independent method for fitting the emission lines in the coadded spectra of the SDSS-RM sample.  \citet{Shen16b} use a slightly different sample than ours, including only 134 objects with both \heii\ and [\oii] due to a combination of spectral-fitting redshift and/or wavelength limits and the reliability of the automated fitting process in identifying and fitting the lines in cases where they are weak.  This may contribute to the relatively smaller systematic shifts and scatter they find compared to the present results.

In addition, \citet{Shen16b} find luminosity trends in relative line shifts in the SDSS-RM sample, or equivalently, redshift determinations based on cross-correlation with different emission lines, consistent with other work \citep[HW10; R11;][]{Shen11, Shen12}.  HW10 discuss the systematic effects in cross-correlation redshift determination between lines as due to SED effects --- for example, the differing line ratios within the \ciii] blend.  These are most certainly related to the EV1 effects we are interested in here. HW10 do not directly take these effects into account, although they do apply a correction as a function of quasar luminosity that is arguably related, but there is much scatter.  We see from the right panel of Figure~\ref{F_otherzcompare} that there is a weak correlation as a function of luminosity for our \heii--[\oii] velocity shift measurements, consistent with the results of \citet{Shen16b}.  However, there is significant scatter about this weak correlation, arguably driven by the lack of scatter toward positive shifts at high L, rather than a consistent trend across the full luminosity range.  Nonetheless, this trend goes in the direction expected from the SED and EV1 effects, where the highest luminosity, highest accretion-rate sources are more likely to have signatures of quasar outflows in their spectra (see also discussions by, e.g., R11 and references therein).

\subsection{{\rm \heii} Redshifts Compared to BOSS Pipeline Redshifts}
\label{S_compare2BOSS}

\begin{figure}
\epsscale{1.2}
\plotone{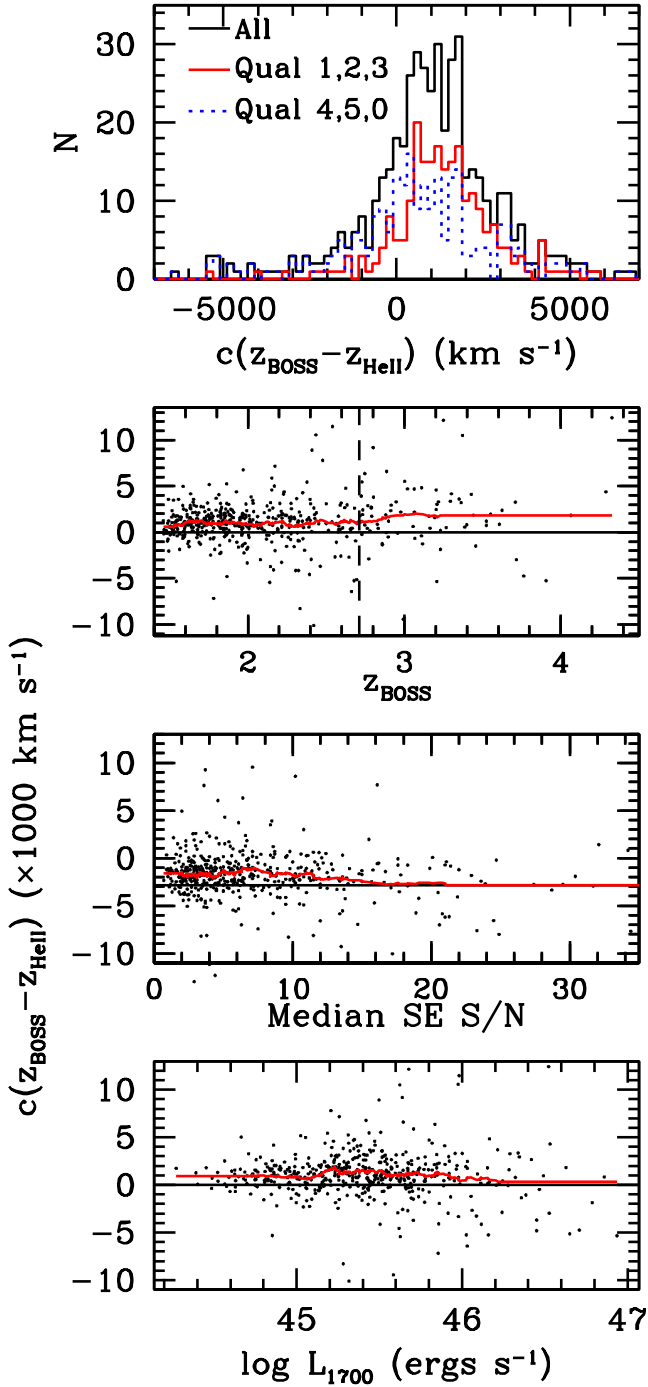}

\caption{The difference between BOSS-pipeline and \heii\ redshifts. The top panel compares the BOSS-pipeline redshifts to those from the \heii\,$\lambda$\,1640 line centroid measured from the coadded spectra; black represents the full sample, and the ``more'' (Q = 1,2,3) and ``less" (Q = 4,5,0) reliable subsamples (see Section \ref{S_Analredshifts}) are shown by the red solid and blue dotted histograms, respectively. The bottom three panels, ordered highest to lowest, show the redshift differences as a function of (i) BOSS redshift, (ii) median single-epoch S/N, and (iii) luminosity. The red solid curves show a 51-pt running mean of the redshift differences, while the horizontal black lines are a reference for equal BOSS and \heii\ redshifts.  The vertical dashed line in the highest of the three panels shows where the \mgii\ emission line redshifts out of the BOSS wavelength range.} 

\label{F_zbiases}
\end{figure}

We investigate the possibility for biases in the BOSS-pipeline redshifts due to quasar diversity using our \heii\ redshifts described in Section \ref{S_Analredshifts}. Table \ref{T_shifts} quantifies the redshift differences we find between BOSS-pipeline and \heii-based redshifts, which are shown in Figure~\ref{F_zbiases}.  We look separately at the full sample as well as the ``more'' (Q = 1,2,3) and ``less" (Q = 4,5,0) reliable subsamples.  The peak of the full sample distribution is broad, but generally, we find a median shift in the BOSS redshifts of 1080 \kms\ relative to those based on \heii, with a 1$\sigma$ (3$\sigma$) uncertainty on the median from 10,000 bootstrap trials of 90\,\kms\ (225\,\kms).  This systematic difference is significantly larger (greater than a factor of three) than the median shift between \heii\ and [\oii] that can be explained by quasar diversity effects in the \heii\ redshifts.  The dependence of this redshift difference on redshift shows a relatively larger bias for $z\gtrsim$\,2.8, compared with lower redshifts, probably because this corresponds to the redshift where \mgii\ exits the BOSS spectral range.  There is also a large bias for systems with low S/N.  This is generally consistent with results of the effects of data quality on measurements of emission line properties \citep{Denney16a}.

The bottom panel of Figure~\ref{F_zbiases} suggests that quasar diversity systematics remain present in the BOSS-pipeline redshifts, despite the improvements we expected by using PCA analysis over strict emission line cross correlation with a composite spectrum.  We find the smallest BOSS--\heii\ redshift differences at the highest luminosities, log $L_{1700}>46$, but extrapolation in Figure~\ref{F_otherzcompare} suggests that these luminosities have the largest deviations between \heii\ and [\oii] --- up to $\sim$2000\,\kms --- presumably owing to EV1 quasar diversity effects.  This seems to indicate that {\it both} the BOSS and the \heii\ redshifts are biased (i.e., underestimating the redshift) at these high luminosities, which is the regime where the EV1 effects make \heii\ a poorer redshift indicator. On the other hand, at lower luminosities, we find differences between \heii\ and BOSS-pipeline redshifts of $\sim$1000\,\kms, or more.  Yet, these lower luminosity sources have \heii\ redshifts consistent with those from [\oii] and are thus likely to be probing the systemic redshift.  

These results suggest that both high-luminosity and low-luminosity quasars are biased in opposite senses:  BOSS redshifts of high luminosity quasars are underestimated, while those of low-luminosity quasars are overestimated. This effectively compresses the redshift--luminosity plane --- or the parameter space of any other measured or physical properties with similar dependencies.  In other words, cross-correlation redshifts have an ``averaging" bias in not accounting for the dependence of quasar line shifts on luminosity (cf. HW10) or some more physical underlying source.  This was in some sense done by design in SDSS-pipeline redshifts, where the \citet{VandenBerk01} template was created with \civ\ at 1546\AA, so that contributing quasars that were noted to have a range in \civ\ peak wavelengths spanned the template wavelength $\pm \sim$1000\,\kms, rather than a template with \civ\ at the expected 1549\AA, with objects showing shifts $\sim$0--2000\,\kms\ (G.~T.~Richards, private communication).  The best explanation we can determine for this effect remaining so large in BOSS-pipeline redshifts is the implicit dependence of the BOSS-pipeline redshifts on cross-correlation redshifts: the PCA training sample redshifts are still based on comparison with a composite spectrum, despite the improvements afforded by PCA.

\subsection{A Comparison to HW10 Redshifts}
\label{S_HWcompare}

\begin{figure*}
\epsscale{1.1}
\plotone{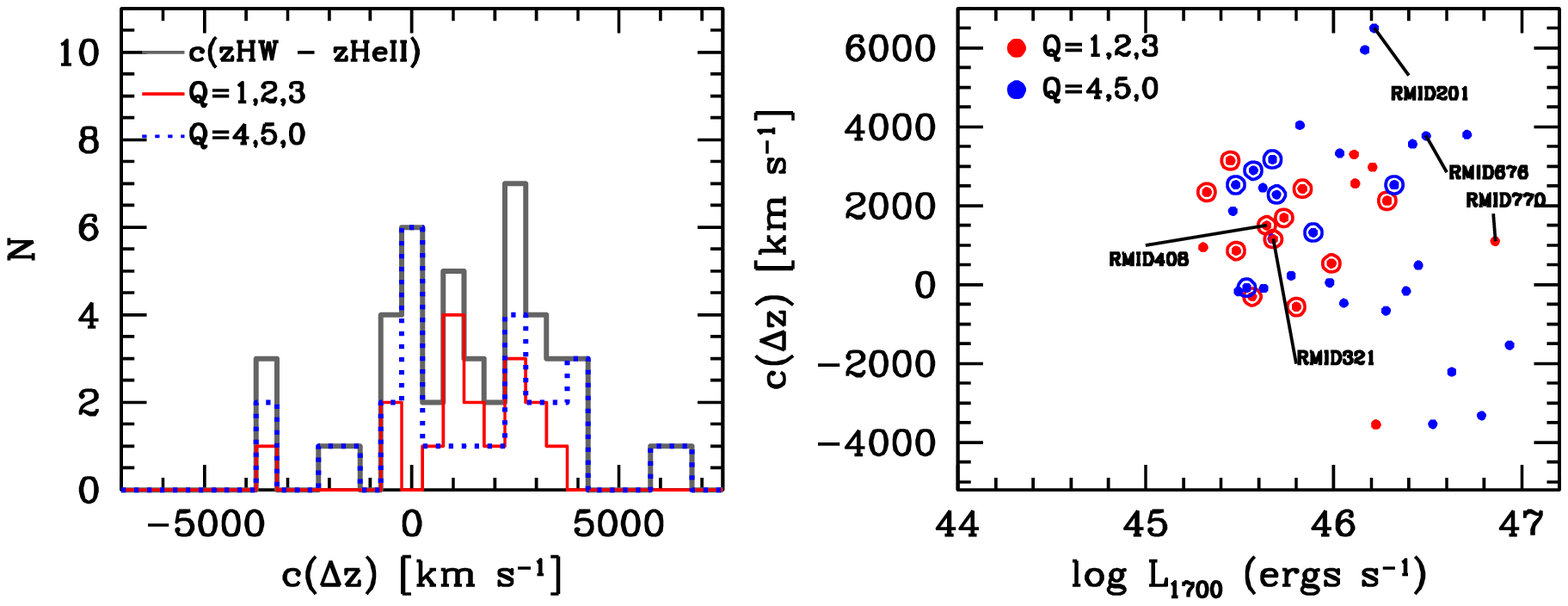}

\caption{{\it Left}: Same as the top panel of Figure~\ref{F_zbiases} but between our \heii\ redshifts and the HW10 redshifts for the shared sample of 47 objects. {\it Right}: Same as the right panel of Figure~\ref{F_otherzcompare} but for redshift differences between our \heii\ redshifts and the HW10 redshifts.  The points surrounded by larger open circles are sources that also have [\oii] measurements.  Selected objects discussed in Section~\ref{S_HWcompare} are individually identified.} 

\label{F_HWzcompare}
\end{figure*}

As discussed previously, HW10 attempted to mitigate quasar-diversity biases in SDSS-pipeline redshifts by applying a luminosity correction.  In Figure~\ref{F_HWzcompare} we compare the \heii\ redshifts to HW10 redshifts\footnote{Available from http://das.sdss.org/va/Hewett\_Wild\_dr7qso\_newz/} for the 47 quasars (17 with Q=1, 2, 3) we have in common with HW10.  The left panel of Figure~\ref{F_HWzcompare} and statistics in Table \ref{T_shifts} show a larger shift between our \heii\ redshifts and HW10 redshifts than the small differences we measured between \heii\ and [\oii]. This is, at first glance, surprising, since HW10 made corrections to the SDSS redshifts for the luminosity-dependent biases.

One interpretation of these differences is that there are still quasar emission-line property diversity systematics in the HW10 redshifts, despite the improvements and corrections they employ.  They do still make use of cross-correlation with template spectra and so may still be susceptible to averaging across the diversity to form their composite spectra.  Alternately, we may simply be ``unlucky'' in our overlap sample. Admittedly, the number statistics of our comparison are small and the scatter is large. From a quasar physics standpoint, however, the SDSS spectroscopic flux limit is shallower than that of BOSS and SDSS-RM. So the overlap with our SDSS-RM sample is biased to bright AGN that probes a different part of quasar diversity parameter space. This implies the distribution of overlap with our SDSS-RM sample is not unbiased.  Indeed, the open circles in the right panel of Figure~\ref{F_HWzcompare} show that the 18 objects in our overlap sample with HW10 that also have [\oii] redshifts are not distributed uniformly and, instead, exhibit larger than average shifts between \heii\ and the systemic redshift (assumed from [\oii] measurements).   

\begin{figure*}
\epsscale{1.2}
\plotone{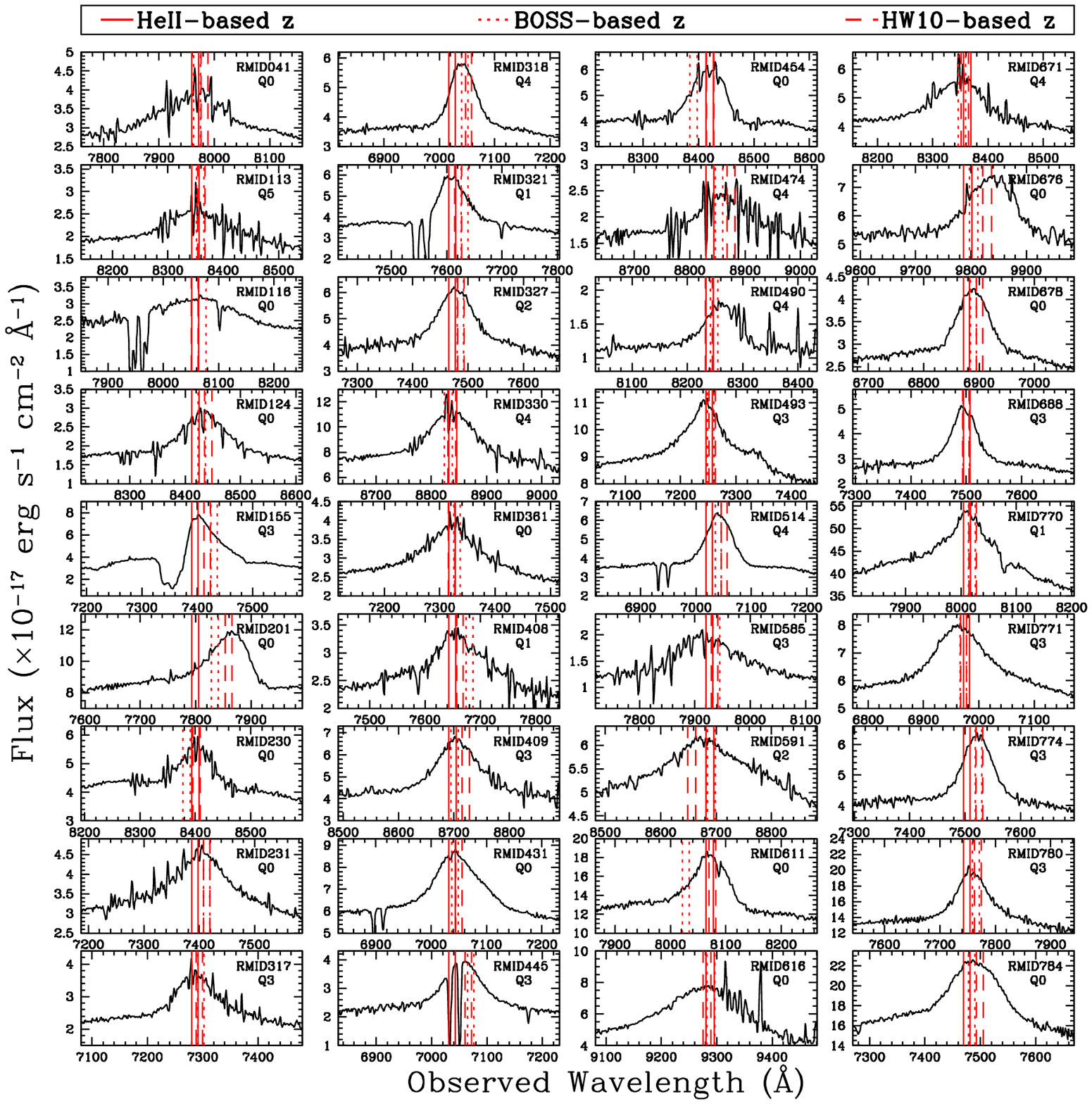}

\caption{\mgii\ emission lines for the 36 SDSS-RM sample objects that also overlap with the HW10 sample.  Vertical lines show the expected location of the \mgii\ doublet peaks given the redshift estimate provided by each method shown in the legend at the top.  RMID and \heii-redshift quality rating are shown in the top right corner of each panel.  All spectra are shown in the observed frame because of the uncertainty of which redshift is correct.} 

\label{MgIIthumbnails}
\end{figure*}

With this biased sample-overlap, we cannot make an objective and comprehensive investigation into the degree to which HW10 redshifts effectively correct for EV1/SED effects in their redshifts, and/or whether they may still be susceptible to quasar diversity biases.  However, we try to gain a small amount of additional insight by looking at the 36 of the 47 objects in the SDSSRM$-$HW10 overlap sample that have \mgii\ present in the spectrum (shown in Figure~\ref{MgIIthumbnails}).  Of these, 30 of the HW10 redshifts are based on \mgii\ cross-correlation, which is generally taken to be the most reliable broad line for determining redshifts for 0.8$< z <$2.8.  For example, within the SDSS-RM sample, \citet{Shen16b} find only a small systematic shift, $-57$\,\kms, on average, between the location of the \mgii\ peak and the \caii\ stellar absorption features, with no luminosity dependence.  There are 15 objects of this subsample that have Q = 1, 2, 3. From inspection of the observed \mgii\ doublet peaks in Figure~\ref{MgIIthumbnails} relative to the predictions from the \heii-based redshift (solid lines), BOSS redshift (dotted lines), and HW10 redshift (dashed lines), there is not an obvious, uniform explanation for the redshift differences.

For the three objects (RMID 321, RMID408, and RMID770) with Q=1 (for which \heii\ redshifts are the most robust), we find that the \heii\ redshift more accurately predicts the peaks of the \mgii\ doublet than the BOSS-pipeline or HW10 methods.  All three of these objects display a shift between the \heii\ and HW10 redshifts of $\sim$1000\kms, consistent with the average \heii--BOSS redshift difference. Yet, RMID321 and RMID408, the two of these three objects that also have [\oii] measurements, show a relative shift between [\oii] and \heii\ of only 11\,\kms\ and $-$52\,\kms, respectively.  At least in these cases, this provides strong evidence that the \heii\ redshift for these objects is trustworthy. 

Consistent with the large scatter in the left panel of Figure 5, there is a lot of inconsistency in the MgII peak-prediction accuracy of the Q=2 and Q=3 HeII redshifts: in some objects, the \heii-based redshift is better (e.g., RMID155), while for others (e.g., RMID774), it makes a much worse prediction than BOSS or HW10.  Sometimes, the HW10 redshift is clearly superior to the BOSS or \heii\ estimate, such as for RMID201 and RMID676, Q=0 cases, where HeII is frequently asymmetric and is often blueshifted with respect to [OII], and/or no discernible narrow \heii\ emission line is present.  These cases highlight the clear improvement in redshifts offered by the luminosity (i.e., quasar diversity) corrections of HW10 over the SDSS- or BOSS- pipeline.

There are two other considerations when comparing results with \mgii, which is likely the best UV broad line to use for redshifts. First, even \mgii\ is sometimes blueshifted with respect to \Hbeta\ and \ob\ \citep[e.g.,][]{Marziani13, Plotkin15}.  Such objects tend to be the same that are argued to have high accretion rates, high luminosity, blueshifts, and low-EW, high-ionization lines \citep[e.g.,][]{Sulentic07, Richards11, Luo15}.  These objects are possibly also related to so-called weak-line quasars that also may have characteristically weak or absorbed X-ray properties and are suggested to lie at the extreme end of this parameter space \citep[e.g.,][]{Luo15, Plotkin15}. Furthermore, \mgii\ likely has little or no emission contribution from the NLR because of the photoionization physics regulating this transition.  As such, diversity in BLR kinematics between objects will affect redshifts based on the \mgii\ peak compared to redshifts determined from narrow emission lines or host-galaxy stellar absorption lines.  This is likely a cause of scatter in the peak shifts between \mgii\ and [\oii] found by \citet{Shen16b} and others.  Second, the \mgii\ doublet {\it ratio} is dependent on the physical conditions of the BLR.  While the doublet is often not fully resolved due to moderate spectral resolution or because of the large BLR velocities, the line peak can still shift among differing physical environments that allow the ratio to vary between 1:1 and 2:1. Interpreting cross-correlation results therefore becomes non-trivial, and such effects will also affect the characteristics of the PCA training set or template spectrum.  Clearly, this is a difficult and multi-faceted problem. While \mgii\ is certainly better than other UV quasar emission lines that exhibit more blending or stronger EV1/quasar diversity effects, \mgii\ will still exhibit diversity due to the physical environment of the nucleus.

\subsection{{\rm \civ} Blueshifts: Implications from BOSS Redshift Biases}

\begin{figure*}
\epsscale{1.2}
\plotone{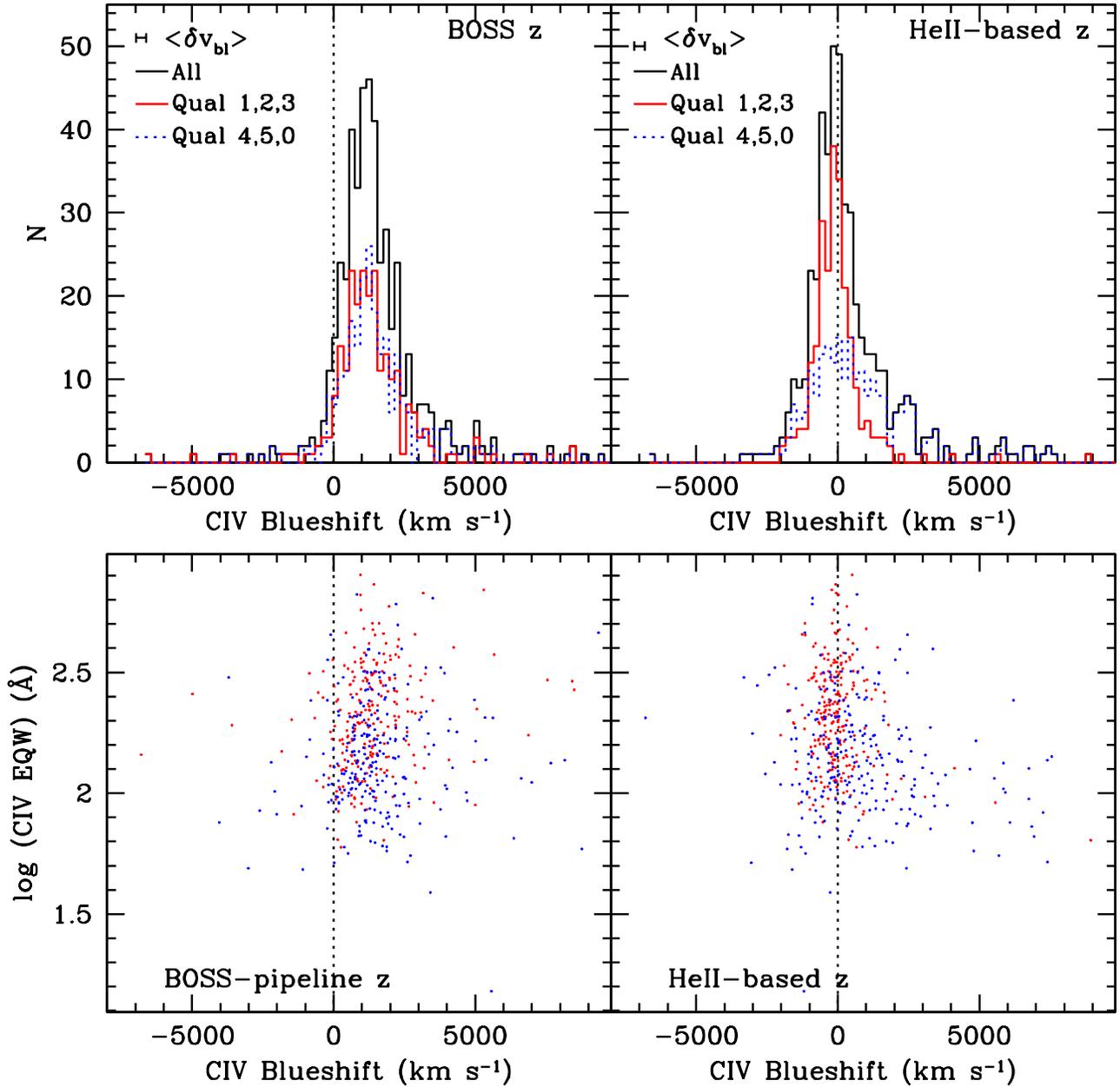}

\caption{\civ\ blueshifts inferred from different redshift determinations. The top panels show implied \civ\ line centroid blueshifts based on the BOSS redshifts (left) and \heii-based redshifts (right).  Colors are the same as the top panel of Figure~\ref{F_zbiases}.  The median \civ\ centroid uncertainty of 183 \kms\ is represented by the $\langle \delta v_{\rm bl} \rangle$ error bar in the top corner.  Here, we use the same convention as R11 that larger blueshifts are indicated by larger positive velocities.  The bottom panels show the same \civ\ blueshifts where the points follow the same color coding as the histograms in the top panels, but are shown with respect to the \civ\ line equivalent width, which are taken from \citet{Denney16a}.} 

\label{F_civblue}
\end{figure*}

Results from our \heii-based redshift analysis suggest that the reliability of current high-$z$ quasar redshifts can be connected to spectroscopic quasar diversity related to the intrinsic, physical differences in the nuclear structure and/or SED, and is likely related to accretion rate \citep[see, e.g.,][]{Leighly04b, Baskin13, Luo15}.  Consequently, many of the \civ\ blueshifts previously inferred for SDSS quasars by R11 and others may be similarly biased if the quasar redshift determinations are unreliable.  We still find evidence that \civ\ exhibits line peak blueshifts (as do other high-ionization broad lines, including \heii), {\it but only in some} quasars.  This is demonstrated in Figure~\ref{F_civblue}, where the top panels show the measured \civ\ centroid blueshifts based on the BOSS-pipeline redshifts (top left) and our \heii-based redshifts (top right).  We find different trends for Q=1, 2, 3 sources as for Q=4, 5, 0 sources.  The distinction between these subsamples, while technically set by our ability to reliably determine a \heii-based redshift, is, as argued above, related to the intrinsic differences in the physical properties of these sources.

Objects with strong (or relatively stronger), narrow \heii\ emission components (typically Q = 1, 2, and 3) do not have systematically blueshifted \civ\ emission, while objects without this narrow component can show very large \civ\ blueshifts that are likely caused by SED effects, such as radiation line-driven outflows \citep[see, e.g., R11;][and references therein for further discussion]{Denney12}.  However, Figure~\ref{F_civblue} and the discussions above provide evidence that BOSS-pipeline-based redshifts are even more biased than redshifts based on \heii, even considering the shortcomings of using this line.  The most likely explanation for these biases is due to the measurement of redshifts from cross-correlation with template spectra that are formed from a sample that covers the large physical property parameter space of quasars.  This complicates analyses that try to understand this parameter space using \civ\ diagnostics and any studies that require reliable redshifts as a means to interpret emission- or absorption-line velocity shifts in quasar spectra \citep[e.g.,][]{Khare14}.

Evidence for the averaging effect discussed in Section~1 --- likely due to the ultimate dependence of the BOSS-pipeline redshifts on an average composite spectrum --- is shown in the top left panel of Figure~\ref{F_civblue}.  The distribution of \civ\ blueshifts for Q = 1, 2, and 3 redshifts is co-spatial with the Q = 4, 5, and 0 distribution, and all objects are driven to an ``average" \civ\ blueshift, likely coincident with that imposed by a bias intrinsic in the redshifts of the PCA training set of quasar spectra.  On the other hand, using the \heii-based redshifts, we still find \civ\ blueshifts (some very large), but the lower limit is consistent with no blueshift, within the precision of our measurements.  Importantly, the overall dynamic range of \civ\ blueshifts is much larger now that the averaging bias has been corrected.

Similar evidence is also seen by investigating the \civ\ blueshift--EW parameter space used by R11 (bottom panels of Figure~\ref{F_civblue}).  R11 argue that this observational parameter space traces physical quasar properties with respect to the prevalence of disk-winds (due to differences in mass accretion rate).  Our results are consistent with this picture, but we argue that the effect of disk winds on \civ\ is not as ubiquitous as that indicated by R11 at a significance level that would statistically bias black hole mass estimates, i.e., with a systematic blueshift larger than the typical velocity width measurement uncertainties.  Despite the use of the improved HW10 redshifts by R11, the most likely explanation for the difference between their \civ\ blueshift results and what we find here is a remaining bias because of the shallower flux-limited SDSS sample compared to what we probe with SDSS-RM.  Our analysis here suggests that because the HW10 correction is suboptimal for the lowest accretion-rate region of quasar parameter space, the results of R11 do not accurately describe \civ\ blueshift trends across the full range of quasar properties.

The left panels of Figure~\ref{F_civblue} show that both subsamples --- those with stronger, narrow \heii\ lines (red) and those without (blue) --- occupy the same part of this parameter space when basing the \civ\ blueshifts on BOSS-pipeline redshifts but become separated when the \civ\ blueshift is measured with respect to the \heii\ redshifts.  This further supports the physical picture that is presented by R11:  quasars with significant \civ\ blueshifts are likely those with strong disk winds.  However, the \civ\ blueshift --- suggestive of a dominant disk wind --- is not ubiquitous among the quasar population; a number of quasars show, on average, {\it no} systematic blueshift within our measurement uncertainties.  These are predominantly objects with strong low-velocity (narrow) \heii\ and \civ\ emission-line cores. 

One argument that the disappearance of a systematic \civ\ blueshift for a portion of our targets when using \heii-based redshifts could simply indicate that \heii\ and \civ\ share similar blueshifts and that the BOSS redshifts are not biased.  However, its important to note that the minimal \civ\ blueshift objects are also those with \heii\ emission clearly attributable to a NLR emission that show little to no velocity shift with respect to [\oii] or \mgii\ (e.g., RMID321 and RMID408), so this interpretation is unlikely for this population of quasars.  The {\it broad} \heii\ component (the only visible component for many objects) shows similar shifts as the other high-ionization lines, so broad \heii\ likely does trace \civ.  As such, the results shown in Figure~\ref{F_civblue} are still not fully unbiased.  Our combined results suggest that instead, given accurate, unbiased redshifts, the bottom right panel of Figure~\ref{F_civblue} would likely show (i) a somewhat broader distribution of Q = 1, 2, 3 sources with a lower limit still at zero but a possible mean blueshift of a couple of hundred \kms\ to account for the difference between the observed median \civ\ blueshift of $-118$\,\kms\ (i.e., a redshifted \civ\ peak; see Table 1) and the observed velocity shifts between \heii\ and [\oii] of 339\,\kms, likely predominantly attributable to the Q=2 and 3 objects with broader ``narrow" \heii\ lines, and (ii) a distribution of Q = 4, 5, 0 quasars with a higher mean blueshift because the \heii\ redshift is also likely biased low for high accretion-rate sources with strong winds and no narrow \heii.

\section{Conclusion}
\label{S_Conclusions}

Detailed studies of quasars have clearly shown that there is significant diversity in the details of quasar spectra.  The most prominent differences, termed the EV1 parameter space \citep{Boroson92}, have been linked to physical differences in the quasar environment, most likely determined by the accretion rate.  One known observational consequence of this diversity is the relative shifting of the peaks of quasar emission lines, argued to be related to the strength of disk winds that can be driven at high accretion rates (see R11 and references therein).  The relation between the observed spectroscopic properties of quasars and the accretion rate leads to correlations of quasar properties with luminosity in flux-limited samples. As a consequence, determining redshifts for quasars is not as straightforward as determining redshifts for galaxies if the desired precision is $\lesssim 10^3$\,\kms. This is because the mean quasar spectrum changes as a function of redshift due to the observed luminosity dependence of the emission-line properties.

In this work, we investigated the degree to which this quasar diversity affects redshift determinations for the sample of $z>1.46$ SDSS-RM quasars by using redshifts based on the \heii\,$\lambda$1640 emission-line centroid.  We are not advocating that \heii\ ultimately be used as a general tool for determining high redshifts.  Its reliability depends on the identification of the narrow component that is only present in some quasars, since the peak wavelength of broad \heii\ is also susceptible to shifts across EV1 parameter space in the absence of a narrow-line component.  Furthermore, \heii\ is often low-EW, and therefore cannot always be cleanly isolated in lower-S/N ``survey-quality" spectra. Nonetheless, it has benefits for estimating high-$z$ quasar redshifts within the constraints on its observed properties that are set by physically-motivated quasar diversity characteristics.

We found that \heii-based redshifts in the SDSS-RM sample are consistent, on average, with [\oii]-based redshifts to within 1$\sigma$ of the PrepSpec line center measurement uncertainties.  Observed shifts larger than these uncertainties are predominately found in the brightest quasars.  [\oii]-based redshifts should not be susceptible to line shifts due to the diversity in quasar physical properties deep within the nucleus, which causes the velocity shifts primarily in the broad and/or high-ionization lines, because it is emitted from gas from much larger radii and/or from elsewhere in the quasar host.  Objects with Q=1 showed a median \heii\ shift with respect to [\oii] of only $-146\,$\kms, which is well within the median statistical uncertainty of the Prepspec [\oii] measurements.  This suggests that the redshifts of high-redshift quasars with strong narrow \heii\ can be reliably determined to this level of precision from this line alone. The distributions of \heii-to-[\oii] velocity shifts for the full sample suggest that \heii\ exhibits, on average, a small blueshift with respect to the systemic quasar rest frame, on the order of a couple hundred \kms, assuming that [\oii] is a better proxy for this.  However, the \heii--[\oii] shifts can be large if the emission-line peak of \heii\ cannot be attributed to emission from the NLR. These shifts are not enough to explain the full bias in the BOSS redshifts indicated by this investigation.

By comparing BOSS-pipeline redshifts to our \heii-based redshifts, we found evidence that the BOSS-pipeline redshifts may be biased, possibly overestimating the redshifts of some high-$z$ quasars by $\sim$1000 \kms, on average (see Fig.\ \ref{F_zbiases}), while underestimating the redshifts of a smaller population because of the imprint of the physical quasar diversity on observed spectroscopic properties.  The main goal of the BOSS survey to measure baryonic acoustic oscillations in the \Lya\ forest of high-$z$ quasars \citep[e.g.,][]{Delubact15} did not depend on quasar redshifts having a precision greater than this limit, so the great success of the BOSS survey in that regard is unaffected by these results.  However, BOSS spectra are now being used for a myriad of other purposes that this apparent bias may impact if studies require redshifts to higher precision that this measured systematic difference.

The likely source of this bias in the SDSS-pipeline redshifts is their basis on a comparison to (or in the case of the BOSS pipeline, from a PCA training set that ultimately has redshifts determined from) composite quasar spectra.  In the creation of such spectra, objects that exhibit systematic emission-line shifts are coadded with those that do not.  This creates an ``averaging'' effect in terms of the location of the peaks of quasar emission lines.  Because the shifts between the lines are dependent on luminosity for flux-limited samples, the degree of bias introduced depends on how closely the observed spectroscopic properties of the sample used for making the quasar composite resembles the sample to which it is applied.

We additionally investigated the impact of these redshift biases on physical diagnostics of quasars based on \civ\ emission-line shifts.  Adopting the \heii-based redshifts for our sample results in a distribution of \civ\ blueshifts that has a larger range than that found for the HW10-corrected SDSS-pipeline redshifts (cf.\ R11).  We find a median for our sample that is statistically consistent with zero blueshift within our measurement uncertainties, for objects with a strong \heii\ narrow component, and an extended tail to large blueshifts, dominated by objects with little or no narrow \heii\ line emission.  This general picture is consistent with other current studies using more sophisticated redshift determination methods for high-$z$ quasars (P.~Hewett, private communication) and with a physical picture where SED-dependent effects can cause a range in \civ\ blueshifts, but that \civ\ blueshifts are not ubiquitous across the quasar population within the precision probed by our study.  

This result has implications for \civ-based BH mass estimates, the reliability of which has been called into question partially because of the presumed ubiquitous presence of these blueshifts and their being attributed to non-virial motions in the BLR.  If instead, \civ\ blueshifts are not ubiquitous and redshifts are reliably determined, carefully chosen samples can largely avoid objects observed to exhibit presumed non-virial BLR motions much larger than the uncertainties in the line width measurements used for estimating the black hole masses, thus leading to reliable black hole mass estimates for such samples based on current calibrations.  Future work (Bizogni et al., in preparation) is aimed at formulating new calibrations for single-epoch \civ-based masses that will account for these large \civ\ blueshifts and better account for other non-virial components, such as the non-variable, low-velocity core \citep[see][]{Denney12}.

Our results suggest that future improvements in quasar template-based redshift determinations could potentially be made by creating multiple templates along the Eigenvector 1 (EV1) sequence of quasars properties that spans the observed spectroscopic diversity of quasars.  Template selection for each quasar can then be made by comparing as many measurable spectroscopic properties as possible that have been found to relate to EV1 properties, e.g., line ratios, emission-line shapes, line EW, luminosity, continuum slope, X-ray soft excess and power-law slope, etc.\ \citep{Boroson92}.  Alternately, for PCA-based redshifts, efforts should be made to formulate a PCA training set composed of quasars covering this same EV1 parameter space, but for which the spectra of each training-set quasar is of high S/N and covers both UV and optical rest wavelengths.  Both requirements are important so that the redshifts of the training set are as robust as possible (e.g., based on multiple, unblended narrow emission lines and/or host-galaxy absorption lines, which are often of low EW and require high-quality data to isolate).

\acknowledgements We are grateful to Yue Shen for producing the coadded SDSS-RM spectra used in this work and to C.~S.~Kochanek for discussions and editorial contributions.  We thank P.~Hewett and G.~T.~Richards for stimulating discussions that improved the content and clarity of this work.  KDD is supported by an NSF AAPF fellowship awarded under NSF grant AST-1302093. WNB acknowledges support from NSF grant AST-1516784. CJG acknowledges support from NSF grant AST- 1517113. BMP is grateful for support from NSF grant AST-10008882. KH acknowledges support from STFC grant ST/M001296/1. LCH thanks Carnegie Observatories for providing telescope access and acknowledges financial support from Peking University, the Kavli Foundation, the Chinese Academy of Science through grant No. XDB09030102 (Emergence of Cosmological Structures) from the Strategic Priority Research Program, and from the National Natural Science Foundation of China through grant No. 11473002. JRT acknowledges support from NASA through Hubble Fellowship grant \#51330 awarded by the Space Telescope Science Institute, which is operated by the Association of Universities for Research in Astronomy, Inc., for NASA under contract NAS 5-26555. Funding for SDSS-III has been provided by the Alfred P. Sloan Foundation, the Participating Institutions, the National Science Foundation, and the U.S. Department of Energy Office of Science. The SDSS-III web site is http://www.sdss3.org/. SDSS-III is managed by the Astrophysical Research Consortium for the Participating Institutions of the SDSS-III Collaboration including the University of Arizona, the Brazilian Participation Group, Brookhaven National Laboratory, University of Cambridge, Carnegie Mellon University, University of Florida, the French Participation Group, the German Participation Group, Harvard University, the Instituto de Astrofisica de Canarias, the Michigan State/Notre Dame/JINA Participation Group, Johns Hopkins University, Lawrence Berkeley National Laboratory, Max Planck Institute for Astrophysics, Max Planck Institute for Extraterrestrial Physics, New Mexico State University, New York University, Ohio State University, Pennsylvania State University, University of Portsmouth, Princeton University, the Spanish Participation Group, University of Tokyo, University of Utah, Vanderbilt University, University of Virginia, University of Washington, and Yale University.




\begin{deluxetable}{lccccccc}
\tablecolumns{8}
\tablewidth{7.0in}
\tablecaption{Quasar Redshift Differences and \civ\ Blueshifts}
\tabletypesize{\scriptsize}
\tablehead{
\colhead{Distribution} &  \colhead{Sub-sample} & \colhead{Number} & \colhead{Median Sample\tablenotemark{a}} & \colhead{Distribution} & \colhead{Distribution} & \colhead{Distribution} & \colhead{Distribution}\\
\colhead{Property}&\colhead{Description} & \colhead{of Obj.} & \colhead{S/N} & \colhead{Median\tablenotemark{b}} & \colhead{HIPR\tablenotemark{b}}&
\colhead{Mean\tablenotemark{b}} &\colhead{Std.\ Dev.\tablenotemark{b}}\\
\colhead{(1)} &
\colhead{(2)} &
\colhead{(3)} &
\colhead{(4)} &
\colhead{(5)} &
\colhead{(6)} &
\colhead{(7)} &
\colhead{(8)} 
}

\startdata
{\bf Redshift Differences:} &&&&&&& \\
BOSS$-$\heii\			& All				& 482	& 25.7	& 1080	& 1605	& 1057	& 2459	\\
BOSS$-$\heii\			& Q = 1,2,3		& 237	& 23		& 1320	& 1230	& 1366	& 1928	\\
BOSS$-$\heii\			& Q = 4,5,0		& 245	& 29.7	& 690	& 2085	& 758	& 2854	\\
BOSS$-$HW10			& HW10+SDSSRM	& 47		& 61.5	& -386	& 1781	& -618	& 2112	\\
BOSS$-$\heii\			& HW10+SDSSRM	& 47		& 61.5  	& 990	& 1755	& 948	& 2782	\\	
HW10$-$\heii\			& HW10+SDSSRM	& 47		& 61.5  	& 1500	& 1898	& 1566	& 2644	\\
HW10$-$\heii; Q=1,2,3	& HW10+SDSSRM	& 17		& 81.1  	& 1500	& 1641	& 1310	& 1687	\\
HW10$-$\heii; Q=4,5,0	& HW10+SDSSRM	& 30		& 59.7  	& 1319	& 2231	& 1711	& 3076	\\
\heii$-$PS\heii\			& All				& 482	& 25.7	& 54		& 475	& 66		& 617	\\
\heii$-$PS\heii\			& Q = 1,2,3		& 237	& 23		& 98		& 402	& 130	& 472	\\
\heii$-$PS\heii\			& Q = 4,5,0		& 245	& 29.7	& 5		& 620	& 3		& 726	\\
\heii$-$PS[\oii]			& SDSSRM w/ [\oii]	& 154	& 24.7	& -348	& 780	& -354	& 775	\\
\heii$-$PS[\oii]; Q=1,2,3	& SDSSRM w/ [\oii]	& 87		& 23.2	& -339	& 643	& -377	& 667	\\
\heii$-$PS[\oii]; Q=4,5,0	& SDSSRM w/ [\oii]	& 67		& 25.1	& -455	& 963	& -325	& 901	\\
\hline \\ [-1.5ex]
{\bf \civ\ Blueshifts:} &&&&&&& \\
BOSS-based $z$		& All				& 482	& 25.7	& 1258	& 664	& 1526	& 1874	\\
BOSS-based $z$		& Q = 1,2,3		& 237	& 23		& 1217	& 615	& 1394	& 1740	\\ 
BOSS-based $z$		& Q = 4,5,0		& 245	& 29.7	& 1301	& 729	& 1652	& 1989	\\
\heii-based $z$			& All				& 482	& 25.7	& 26		& 706	& 469	& 1899	\\
\heii-based $z$			& Q = 1,2,3		& 237	& 23		& -118	& 362	& 29		& 1323	\\
\heii-based $z$			& Q = 4,5,0		& 245	& 29.7	&  459	& 1111	& 895	& 2247	
\enddata

\tablenotetext{a}{The Median S/N is based on the distributions of Coadded spectra.  The S/N is measured per Angstrom, integrated over an emission-line-free continuum window, $\Delta W$, covering many resolution elements near restframe 1700\AA.}
\tablenotetext{b}{The median, HIPR, mean, and standard deviation (Std.\ Dev.) values are in units of \kms.}

\label{T_shifts}
\end{deluxetable}



\end{document}